\documentclass[aps,prl,twocolumn, groupedaddress, superscriptaddress, floatfix, longbibliography]{revtex4-1}

\usepackage{amsmath}
\usepackage{amssymb}
\usepackage{graphicx}
\usepackage{float}
\usepackage[english]{babel}
\usepackage{dsfont}
\usepackage{epstopdf}
\usepackage{soul}
\usepackage{color}
\usepackage{braket}
\usepackage[normalem]{ulem}

\usepackage[colorlinks=true,
            linkcolor=blue,
            urlcolor=blue,
            citecolor=blue]{hyperref}

\definecolor{SMblue}{rgb}{0,0,0}

\emergencystretch=\maxdimen
\usepackage{titlesec}
\titlespacing\subsection{0pt}{12pt plus 4pt minus 2pt}{12pt plus 2pt minus 2pt}
\titlespacing\section{0pt}{12pt plus 4pt minus 2pt}{7pt plus 2pt minus 2pt}
\titlespacing\subsection{0pt}{12pt plus 4pt minus 2pt}{7pt plus 2pt minus 2pt}

\AtBeginDocument{%
    \newwrite\bibnotes
    \def\bibnotesext{Notes.bib}
    \immediate\openout\bibnotes=\jobname\bibnotesext
    \immediate\write\bibnotes{@CONTROL{REVTEX41Control}}
    \immediate\write\bibnotes{@CONTROL{%
    apsrev41Control,author="08",editor="1",pages="1",title="0",year="1"}}
     \if@filesw
     \immediate\write\@auxout{\string\citation{apsrev41Control}}%
    \fi
}%

\begin{document}
\title{Period-doubled Floquet Solitons}
\author{Sebabrata~Mukherjee}
\email{mukherjee@iisc.ac.in}
\affiliation{Department of Physics, Indian Institute of Science, Bangalore 560012, India}
\affiliation{Department of Physics, The Pennsylvania State University, University Park, PA 16802, USA}
\author{Mikael C.~Rechtsman}
\email{mcrworld@psu.edu}
\affiliation{Department of Physics, The Pennsylvania State University, University Park, PA 16802, USA}
\date{\today}

\begin{abstract}
We propose and experimentally demonstrate a family of Floquet solitons in the bulk of a photonic topological insulator that have double the period of the drive.  Our experimental system consists of a periodically-modulated honeycomb lattice of optical waveguides fabricated by femtosecond laser writing. We employ a Kerr nonlinearity in which self-focusing gives rise to spatial lattice solitons. Our photonic system constitutes a powerful platform where 
the interplay of time-periodic driving, topology and nonlinearity
can be probed in a highly tunable way.\\

\vspace{-1mm}
\noindent
\end{abstract}

\maketitle

{\it Introduction.$-$} Driving a system periodically in time, known as Floquet engineering~\cite{goldman2014periodically, eckardt2017colloquium, weitenberg2021tailoring}, is a powerful method of creating novel Hamiltonians and topologically nontrivial materials~\cite{klitzing1980new, thouless1982quantized, raghu2008analogs, lu2014topological, ozawa2019topological}.
It is useful for predicting and realizing novel phases of matter that are sometimes inaccessible in static systems.
The idea of creating artificial quantum systems and controlling their features by external driving
has been gaining enormous interest in recent years
because of the development in the tunability in many experimental platforms, such as solid-state devices, photonics and ultra-cold atoms in optical lattices. 
In these experiments, various driving protocols are implemented, and the interplay with disorder and nonlinearity/interactions may be directly studied.
Periodic driving has indeed been exploited for exploring the dynamics of noninteracting~\cite{dunlap1986dynamic, longhi2006observation, lignier2007dynamical, szameit2010observation, mukherjee2015modulation}, as well as many-body~\cite{zenesini2009coherent} lattice systems, artificial magnetic fields~\cite{aidelsburger2011experimental, mukherjee2018experimental}, topological bands
~\cite{rechtsman2013photonic, jotzu2014experimental, aidelsburger2015measuring}, the spontaneous breaking of time translation symmetry \cite{wilczek2012quantum, choi2017observation, zhang2017observation} and also for studying nonequilibrium physics~\cite{reitter2017interaction, rubio2020floquet}. 
%


Photonic systems made of evanescently-coupled waveguide arrays are an established platform for exploring 
a wide range of intriguing phenomena~\cite{garanovich2012light}, including those originating from time-periodic driving, non-trivial topology and nonlinearity. 
The capability of imaging optical fields along the propagation direction ($z$) is beneficial for accessing the effective time ($t \leftrightarrow z$) evolution of an initial state. In other words, the distance along the propagation axis substitutes for the time coordinate, and importantly, a suitable driving can effectively break the `time'-reversal symmetry creating topologically non-trivial quasienergy bands. In such media, nonlinear effects arise naturally at high optical intensities via the optical nonlinearity available in the ambient medium - for which the Kerr effect is the relevant one in this experiment.
%
Some recent examples of nonlinear topological effects include 
the formation of Floquet solitons in a topological bandgap~\cite{lumer2013self, mukherjee2020observation, parker2021floquet}, topological edge solitons~\cite{ablowitz2014linear,leykam2016edge, mukherjee2021observation}, 
nonlinearity-induced {\it local} topological edge states~\cite{maczewsky2020nonlinearity},
and nonlinear Thouless pumping~\cite{jurgensen2021quantized, jurgensen2022quantized}. 
In this context, we also highlight recent works on 
interacting Floquet polaritons~\cite{clark2019interacting},
modulation-induced nonlinearity in magnetic insulators~\cite{shan2021giant} and optical devices~\cite{goldman2022floquet}. 

Spatial solitons are localized shape-preserving waves formed when nonlinearity balances linear diffraction of a wave packet~\cite{barthelemy1985propagation, christodoulides1988discrete, segev1992spatial, eisenberg1998discrete}. 
Floquet solitons are nonlinear waves in the Floquet sense: Due to the periodic nature of the system's Hamiltonian, they reproduce themselves with each driving period, up to a phase factor~\cite{lumer2013self, mukherjee2020observation, parker2021floquet}.
Here we propose and observe a new type of Floquet solitons whose periodicity is twice the driving period. 
In other words, the spatial intensity profile of these period-doubled solitons exhibits a periodic motion with a period of two Floquet cycles.
Note that the period-doubled solitons are unique to nonlinear systems since the linear Floquet modes always repeat themselves after each complete driving period during propagation. 
%
%
Our experimental system consists of laser-fabricated anomalous Floquet topological insulators~\cite{mukherjee2017experimental, maczewsky2017observation, wintersperger2020realization} where light waves exhibit robust edge transport, even though the standard topological invariants (e.g., the Chern number) of all quasienergy bands are zero -- hence the name `anomalous'~\cite{kitagawa2010topological, rudner2013anomalous}.
In our experiments, a self-focusing nonlinearity arises at high optical intensities due to the Kerr effect -- an intensity-dependent modification of the local refractive index induced by nonlinear dielectric polarization.
The nontrivial topology of the photonic lattice is confirmed by probing backscatter-immune unidirectional linear topological edge states. In the nonlinear experiments, we probe the period-doubled soliton family and show the dynamics of the mid-gap soliton. It should be highlighted that the period-doubled solitons coexist with the traditional single-period soliton family~\cite{lumer2013self, mukherjee2020observation} in the same periodically driven lattice. 
Both soliton families exhibit a cyclotron-type rotational micromotion.

\begin{figure}[t!]
\center
\includegraphics[width=\linewidth]{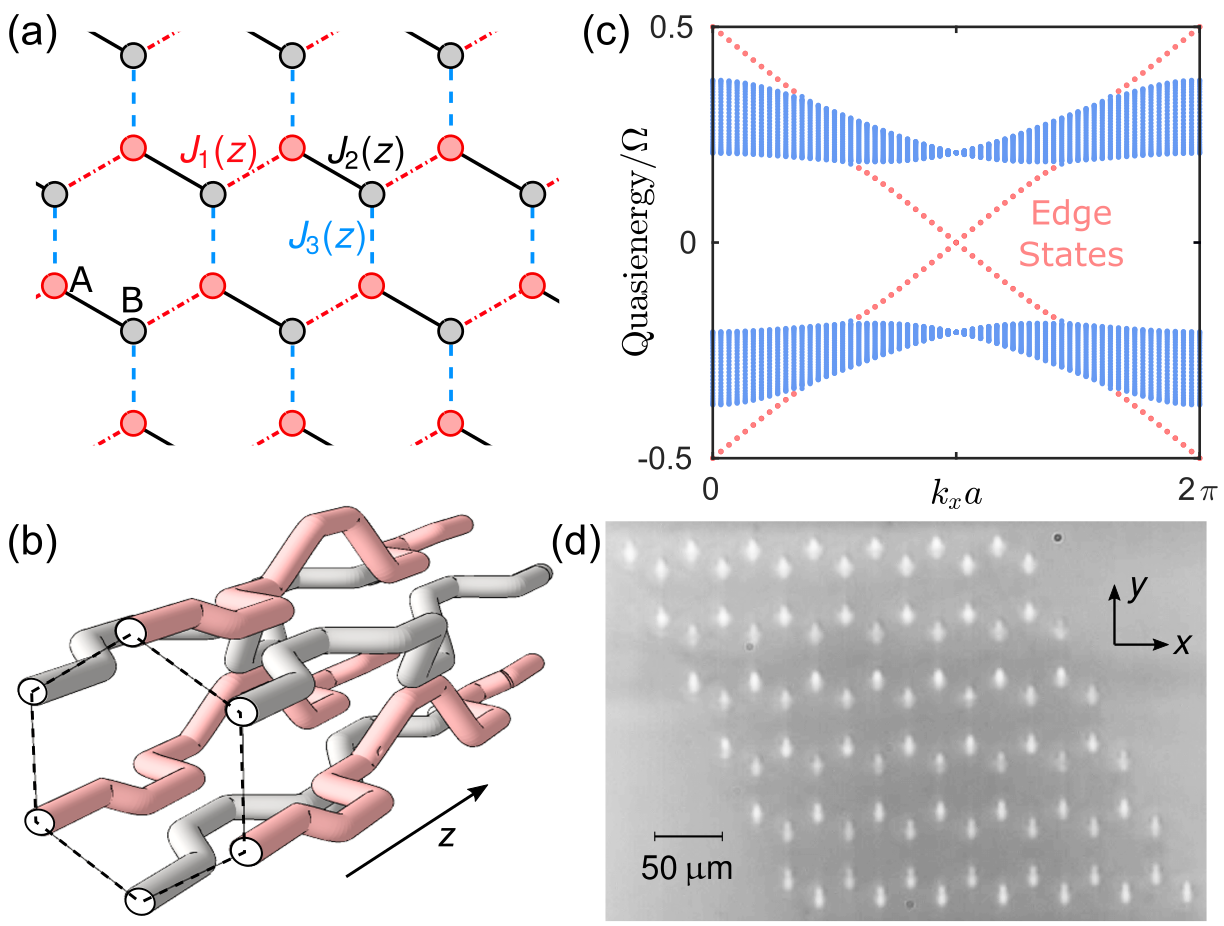}
\caption{(a) A Floquet topological insulator consisting of a honeycomb lattice with only nearest-neighbor couplings $J_{1-3}$ 
that are switched on and off periodically in a three-step clockwise manner. (b) Photonic implementation of (a) using periodically modulated optical waveguides. The propagation distance is denoted by $z$. (c) Quasienergy spectrum for a strip geometry that is periodic along the horizontal $x$-direction. Notice the bulk bands (blue) and topological chiral edge states (red). (d) White light transmission micrograph of the output facet of a photonic honeycomb lattice fabricated using femtosecond laser writing. 
}
\label{fig1}
\end{figure}

{\it Model.$-$} We consider a honeycomb lattice with nearest-neighbor couplings -- each site has three neighbors, and the couplings among them are denoted by $J_{1,2,3}$, see Fig.~\ref{fig1}(a). These couplings are switched on and off periodically in a three-step clockwise manner such that each site is coupled to only one of its neighbors at a given instant. In other words, only the $J_m$ couplings are switched on during the $m$-th driving step where $m\!=\!\{1,2,3\}$. In the absence of nonlinearity, the transfer of light through a specific coupling can be characterized by the parameters $\Lambda_m\!=\! \int {\text{d}}z J_m(z)$. 
The linear tight-binding Hamiltonian of the honeycomb lattice repeats periodically in $z$, $\hat{H}(z+z_0)\!=\!\hat{H}(z)$ with a driving period $z_0\!=\!2\pi/\Omega$. 
In this case, according to the Floquet theorem, the solution of the time-dependent discrete Schr{\"o}dinger equation can be written as $\Phi(z)\!=\! \exp(-i\varepsilon z)\phi(z)$, where $\phi(z)$ and $\varepsilon$ are formally known as the Floquet state and quasienergy. Here, the Floquet state is $z$-periodic, i.e., $\phi(z+z_0)\!=\!\phi(z)$. In the nonlinear case, we show that period-doubled Floquet states can be observed under certain conditions.


The quasienergy spectrum of the $z$-periodic lattice is obtained by diagonalizing the propagator $\hat{U}(z_0)$ over one period. The propagator is defined as
%
$\hat{U}(z_0) \equiv \mathcal{T} \exp \big[ -i \int_0^{z_0}  {\text{d}} z' \hat{H}(z')  \, \big]$, 
where $ \mathcal{T}$ indicates time ordering in $z$. 
Fig~\ref{fig1}(c) presents the quasienergy spectrum for the experimentally realized values of $\Lambda_m$ (see later) -- the bulk (edge) modes are shown in blue (red). The  spectrum was numerically calculated by considering a strip geometry aligned along the vertical direction and periodic along the horizontal direction. 
The bulk quasienergy bands are weakly dispersive, and topological edge states reside on both quasienergy gaps centered around $\varepsilon\!=\!0$ and $\pm \Omega/2$, respectively. The edge modes around $\varepsilon\!=\! \pm \Omega/2$ can cross the bandgap, connecting the two bands. One band of chiral edge modes, propagating in the same direction on a given edge,
exists above and below the bulk bands. In other words, despite the existence of the topological edge modes, the Chern number of each bulk band is zero. The topology of such a driven system, known as anomalous Floquet topological insulator, is captured by the winding number~\cite{rudner2013anomalous} that is unity for both quasienergy gaps. 
In our experiments, this model is implemented using a periodically modulated honeycomb waveguide array, Figs.~\ref{fig1}(b).
Each lattice site is a single-mode optical waveguide near the operational wavelength of $1030$ nm. The waveguide paths are spatially modulated in $x$-$y$-$z$ to obtain the desired periodic evanescent couplings; see also supplementary information~\cite{Suppmat}. 


In the scalar-paraxial approximation, the nonlinear dynamics of light waves through the waveguide lattice can, in general, be governed by the discrete nonlinear Schr{\"o}dinger equation
\begin{eqnarray}
\label{nlse}
i\frac{\partial}{\partial z} \psi_s(z)=\sum_{\left\langle s' \right\rangle} H_{ss'}^{\text{lin}} \psi_{s'} - |\psi_s|^2 \psi_s \; , \label{eq1}
\end{eqnarray} 
where the propagation distance $z$ is analogous to time, $s$ labels the lattice sites, and $H^{\text{lin}}$ is the linear tight-binding Hamiltonian. 
The nonlinear term in Eq.~\ref{nlse} arises due to self-focusing Kerr nonlinearity and is negligible at sufficiently low optical power. 
Note that the coefficient of the nonlinearity term has been set to unity by renormalizing the wave function.
This means the wave function $\psi_s$ at the $s$-th lattice site is determined by the optical power and nonlinearity.  
%
%
The strength of nonlinearity is quantified by the renormalized power $P\equiv \sum_s |\psi_s|^2$ which is conserved
in the absence of optical losses, such as propagation and bend losses. Note that Eq.~\ref{nlse} is mathematically equivalent to the time-dependent Gross-Pitaevskii equation, meaning that our results are applicable to a wide range of bosonic systems with mean-field interactions.

\begin{figure*}[t!]
\center
\includegraphics[width=\linewidth]{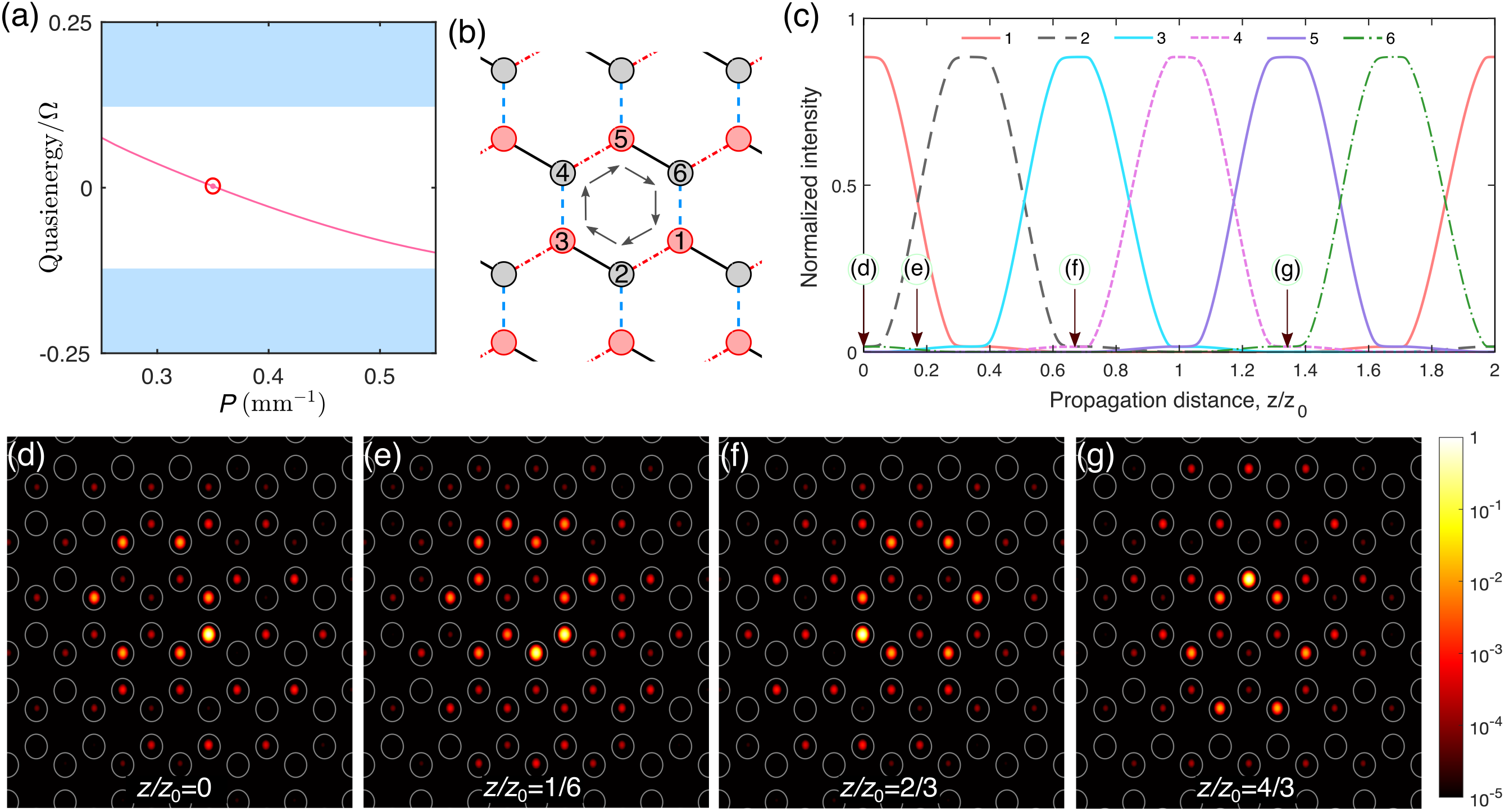}
\caption{(a) Quasienergy spectrum as a function of renormalized power $P$ 
showing bulk modes (blue) and a family of period-doubled solitons (red line). Here, $\Lambda\!=\!0.42\pi$. 
(b) During propagation, the intensity peak of such soliton rotates in a clockwise manner on the six sites (1 to 6) -- the soliton comes back to itself after two driving periods (six driving steps).
(c) Variation of normalized intensity at the six sites showing the dynamics of a period-doubled soliton formed at renormalized power ${P}\!=\!0.35 \Omega$ [indicated by the red circle in (a)]. 
(d-g) Intensity distributions of the period-doubled soliton for $z/z_0\!=\!0, 1/6, 2/3$ and $4/3$, [indicated by arrows in (c)] respectively. Note that the soliton exhibits cyclotron-type clockwise rotation. The white circles indicating the lattice sites are a guide to the eye.
}
\label{fig2}
\end{figure*}

{\it Period-doubled solitons.$-$} Using a self-consistency algorithm~\cite{lumer2013self}, we look for a family of period-doubled Floquet solitions in the bulk of our driven honeycomb lattice. 
It is known that a usual Floquet soliton~\cite{lumer2013self, mukherjee2020observation} exhibits micromotion within the Floquet cycle; however, it reproduces itself (up to a phase factor) periodically after each driving period.
Like a usual Floquet solition, a period-doubled Floquet solition is a nonlinear solution of Eq.~\ref{eq1}; however, it reproduces itself after an integer multiple of two (instead of one) driving periods.
%
%
Since the soliton breaks the spatial periodicity of the lattice, the total Hamiltonian (when we think of the nonlinearity as an induced potential) is now $z$-periodic with a periodicity of $2z_0$. 
This means the quasienergy of the total Hamiltonian is folded, i.e., its span is $[\Omega/4, -\Omega/4]$, which is half of the linear case.
Note that we are dealing with the so-called `Floquet gap solitons' -- i.e., the quasienergy of the soliton resides in a bandgap. Hence, it is evident that an important criterion to find such period-doubled Floquet solitons is the existence of a bandgap after band folding in quasienergy. 
In fact, for obtaining the linear quasienergy spectrum in Fig.~\ref{fig1}(c), the system parameters are chosen to satisfy this criterion.

Fig.~\ref{fig2}(a) presents the quasienergy spectrum as a function of the renormalized power -- the red curve in the bandgap indicates a family of period-doubled solitons bifurcating from the upper linear band (blue). 
Here the red circle indicates the quasienergy of a soliton whose dynamics and intensity pattern are shown in Figs.~\ref{fig2}(d-g).
Each soliton in this family is initially strongly peaked at a single lattice site, and the most intense site rotates clockwise on the six sites [1-6] during propagation, see Fig.~\ref{fig2}(b). Fig.~\ref{fig2}(c) presents the nonlinear dynamics of the normalized optical intensities on the six sites of a period-doubled soliton, whose quasienergy is indicated by the red circle in Fig.~\ref{fig2}(a). Figs.~\ref{fig2}(d-g) show the spatial intensity distributions of the soliton at propagation distances indicated by arrows in Fig.~\ref{fig2}(c).
%
\begin{figure*}[t!]
\center
\includegraphics[width=\linewidth]{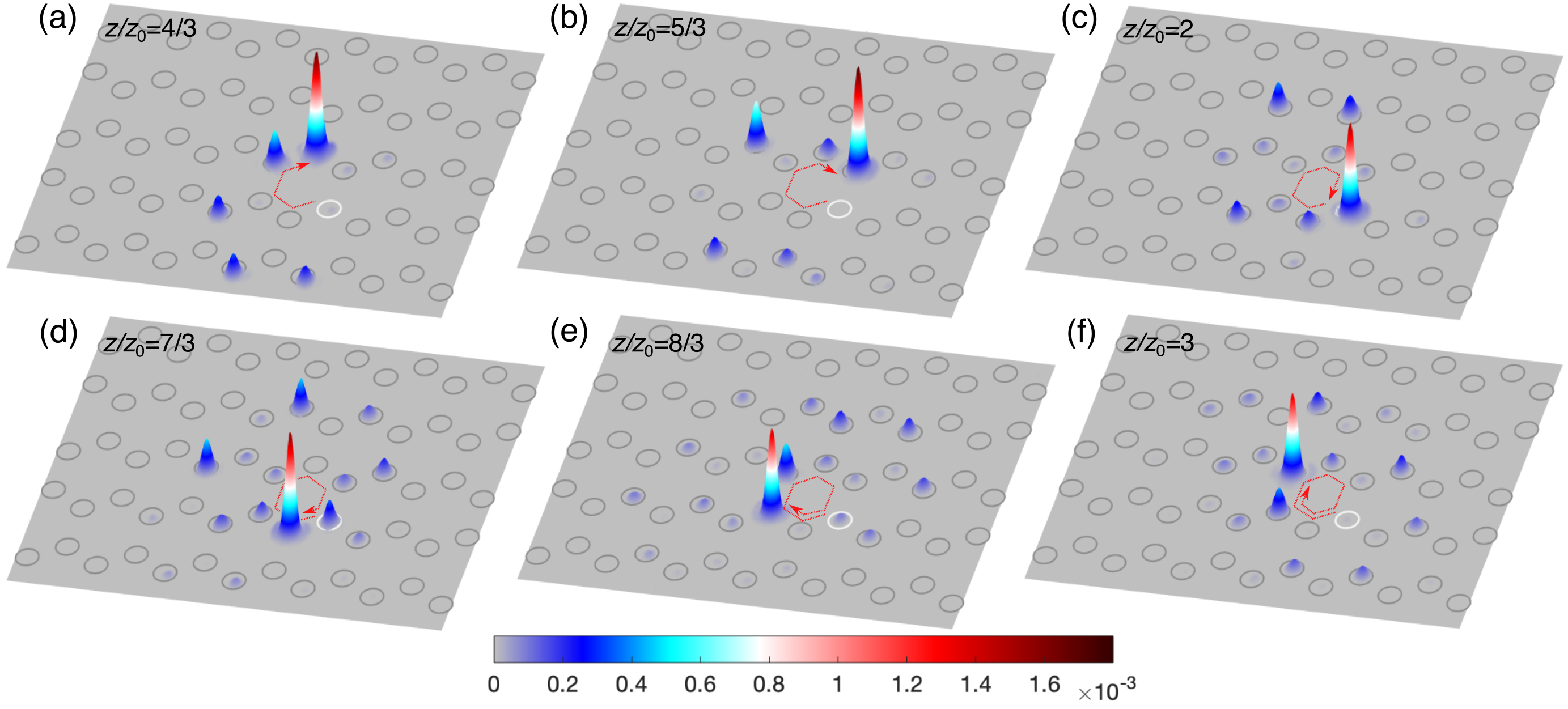}
\caption{Experimentally measured dynamics of a period-doubled soliton in a topological bandgap. Output intensity distributions at six different propagation distances indicated in each image. The input average laser power for all measurements was $P_{\text{in}}\!=\!2.7$ mW. The dark circles are a guide to the eye indicating the lattice sites.
The white circle in each image indicates the site where the light was launched at the input. The red arrow highlights the motion of the soliton peak. 
}
\label{fig3}
\end{figure*}

In addition to the `time'-periodic motion, Floquet solitons show fundamentally different localization features compared to fundamental solitons in static systems. Standard solitons in a static system (e.g., an undriven square lattice) usually become more and more localized as the strength of the nonlinearity increases. 
On the other hand, the spatial extent of the Floquet solitons in Fig.~\ref{fig2}(a) first decreases as a function of nonlinearity, showing maximal localization near the mid-gap quasienergy $\varepsilon \! = \! 0$. The solitons then become delocalized again as nonlinearity is further increased. This is consistent with the fact that solitons near the linear bands are delocalized and become increasingly localized as their (quasi)energy moves away from the linear band. In our nonlinear characterizations, we demonstrate the above-mentioned delocalization to localization to delocalization to prove the existence of the soliton family.

{\it Experiments.$-$} By focusing femtosecond ($260$ fs) laser pulses inside a borosilicate (Eagle XG) substrate, we fabricated optical waveguides with three-dimensional paths, see supplementary information~\cite{Suppmat}.  
We created the driven honeycomb lattice of 86 modulated waveguides using this technique. A white light transmission micrograph of the output facet of the photonic lattice is shown in Fig.~\ref{fig1}(d). 
As discussed in the supplementary information~\cite{Suppmat}, the driving parameter $\Lambda_m$s were estimated to be $1.32 \pm 0.05$. To demonstrate the topological nature of the lattice, we probe edge transport in the linear domain (i.e., $P\! \rightarrow\!0$).
For linear characterization, we couple low power light at a central wavelength of 1030 nm into a single waveguide on the edge and then measure the output intensity pattern after a propagation of three driving periods, $3z_0\!=\!76$ mm. The single site excitation overlaps efficiently $(62\, \%)$ with the edge modes of the lattice. Figs.~\ref{LinearEdgeStates}(a-d) in the supplementary information show the output intensity patterns for four different input excitations indicated by the white circles. In this baseline experiment, we observe the unidirectional linear topological edge modes traveling around a corner (that acts as a defect) without backscattering.

To introduce optical Kerr nonlinearity, we launch laser pulse trains into the desired waveguide of the periodically modulated honeycomb lattice. The strength of the nonlinearity, i.e.,  $P\equiv \sum_s |\psi_s|^2$, is experimentally tuned by varying the laser pulse energy.  The renormalized power and the average laser power are related as follows: $P\equiv \sum_s |\psi_s|^2$ at $z\!=\!0$ was estimated to be $0.08$ mm$^{-1}$ per unit average power in mW. 
It should be noted that the temporal shape of the input state can cause unwanted effects, such as self-phase modulation (i.e., nonlinearity-induced generation of new wavelengths) and chromatic dispersion, which may invalidate Eq.~\ref{eq1}. To minimize these effects, we use temporally stretched and down-chirped pulse trains of $2$ ps pulse width and $5$ kHz repetition rate; see more details in the supplementary information section~\cite{Suppmat}. 

As indicated in Fig.~\ref{fig2}, the period-doubled solitons are strongly peaked at a single lattice site, meaning that the soliton family can be probed by single site input excitation. Laser pulses were efficiently coupled into a single bulk lattice site away from the boundaries. We first measure intensity patterns after a propagation of three driving periods as a function of optical power. In this experiment, output intensity patterns show that the soliton first starts to become more localized, and then delocalizes as a function of optical power -- this behavior is consistent with our numerical prediction, see Figs.~\ref{iteration} and the \ref{IPR} in the supplementary information. To quantify this behavior, we calculate the inverse participation ratio [${\text {IPR}}\!=\!\sum |\psi_s|^4/(\sum |\psi_s|^2)^2$] from the measured intensity patterns. The IPR peak associated with the Floquet soliton near the mid-gap quasienergy $\varepsilon\!=\!0$, was observed at an average laser power of $2.7$~mW. 

We measured the output state at six different propagation distances, $z\!=\![4/3, 5/3, 2, 7/3, 8/3, 3]z_0$, as shown in Fig.~\ref{fig3}. The white circle in each image indicates the site [i.e., site-$1$ in Fig.~\ref{fig2}(b)] where the light was launched at the input, and the red arrow highlights the motion of the soliton peak. After a propagation of $z\!=\!2z_0$, the intensity peak recycles to where the light was injected, see Fig.~\ref{fig3}(c). 
Additionally, note that the intensity peak is at site $4$ [i.e., directly across site-$1$ where the initial state was launched] after a propagation $z\!=\!3z_0$, Figs.~\ref{fig3}(f). 
These observations demonstrate the clockwise rotational motion of the period-doubled soliton near the mid-gap quasienergy $\varepsilon\!=\!0$. 
Due to the temporal shape of the optical pulses, the localization peak is quantitatively lower than what is expected from our numerical calculations. Indeed, the front and rear tails of the pulse exhibit linear dynamics, causing a lower IPR (see Fig.~\ref{IPR}) and some excess diffraction in Fig.~\ref{fig3}. However, the observed peak of IPR along with the nonlinear evolution shown in Fig.~\ref{fig3}, clearly proves the existence of a family of Floquet solitons whose periodicity is twice of the driving cycle.
Although we have only elaborated upon period-doubled solitons in the main text, the same lattice also supports traditional single period Floquet solitons as discussed in the supplementary text~\cite{Suppmat} (See Ref.~\cite{mukherjee2020observation} for the experimental demonstration of single period Floquet solitons.) In the supplementary information, we also show the existence of period-doubled solitons in a sinusoidally-modulated SSH lattice.


{\it Conclusions.$-$} In conclusion, we have proposed and experimentally demonstrated nonlinear Floquet solitons with a periodicity of two driving cycles. These period-doubled Floquet solitons reside in a topological bandgap exhibiting a cyclotron-type rotational motion. This effect clearly represents a non-trivial interplay of periodic driving and nonlinearity. These results are also related to the spontaneous breaking of time translation symmetry observed parametric oscillation, in so-called `time crystals'~\cite{wilczek2012quantum, choi2017observation, zhang2017observation}, 
and period-doubled nonlinear systems~\cite{strogatz2018nonlinear}.  Future directions of research include the investigation the effects of off-site nonlinear terms that are induced by the drive; Floquet solitons in driven systems in the quantum (i.e., low-photon number) limit, where low-energy excitations and associated fluctuations become important; as well as more generally leveraging topological physics to protect nonlinear devices against parasitic effects of fabrication-induced disorder. 

{\it Acknowledgments.$-$}  S.M. and M.C.R. acknowledge support from the Office of Naval Research under Grant Nos. N00014-18-1-2595 and N00014-20-1-2325, and M.C.R. acknowledges the Packard Foundation under Fellowship No. 2017-66821. We thank N. Smith from Corning Glass for providing high-quality Eagle XG glass wafers. S.M. gratefully acknowledges support from IISc and ISRO.

%

\clearpage

\onecolumngrid
\appendix


\newcommand{\beginsupplement}{%
        \setcounter{equation}{0}
        \renewcommand{\theequation}{S\arabic{equation}}%
        \setcounter{figure}{0}
        \renewcommand{\thefigure}{S\arabic{figure}}%
         }
 
 \beginsupplement


\section*{\large {Supplementary Information} \\
Period-doubled Floquet Solitons\\
\small{Sebabrata~Mukherjee and Mikael C.~Rechtsman}
\vspace*{0.3cm}}

\twocolumngrid

In the following sections of the supplementary material, we present experimental and numerical results for completeness and to support the discussion in the main text. Specifically, we discuss photonic device fabrication and characterization process, numerical methods of calculating Floquet solitons and related details.
Additionally, we show the existence of period-doubled solitions in a driven SSH lattice.

{\it Photonic lattice inscription.$-$} The photonic honeycomb lattices were created using femtosecond laser writing~\cite{szameit2010discrete, mukherjee2020observation} -- a powerful technique for fabricating three-dimensional photonic devices.
For this purpose, optical pulse trains of $260$ fs pulse duration, $500$ kHz repetition rate at $1030$ nm central wavelength were generated from a commercial fiber laser system (Menlo BlueCut).
Such laser pulses, focused inside a borosilicate glass substrate (Corning Eagle XG), can locally modify the refractive index ($\sim 10^{-4} - 10^{-3}$) of the substrate. 
The glass substrate was translated once through the focus of the laser beam for fabricating each waveguide. Computer-controlled stable translation was achieved using Aerotech high precision $x$-$y$-$z$ linear stages.
We optimized laser-writing parameters, such as writing speed, pulse energy, numerical aperture of the focusing lens, etc., to produce low-loss single-mode waveguides at the operational wavelength of $1030 \pm 4$ nm. We note that all experiments were performed using the vertical polarization state, and the rotation of polarization (i.e., birefringence) was not detected.
%
Straight waveguides exhibited $0.45$ dB/cm propagation loss. The total insertion (propagation+bend+coupling) loss of our $76$~mm-long modulated waveguides forming the topological photonic lattice was measured to be $4.8$ dB.

\begin{figure}[]
\center
\includegraphics[width=\linewidth]{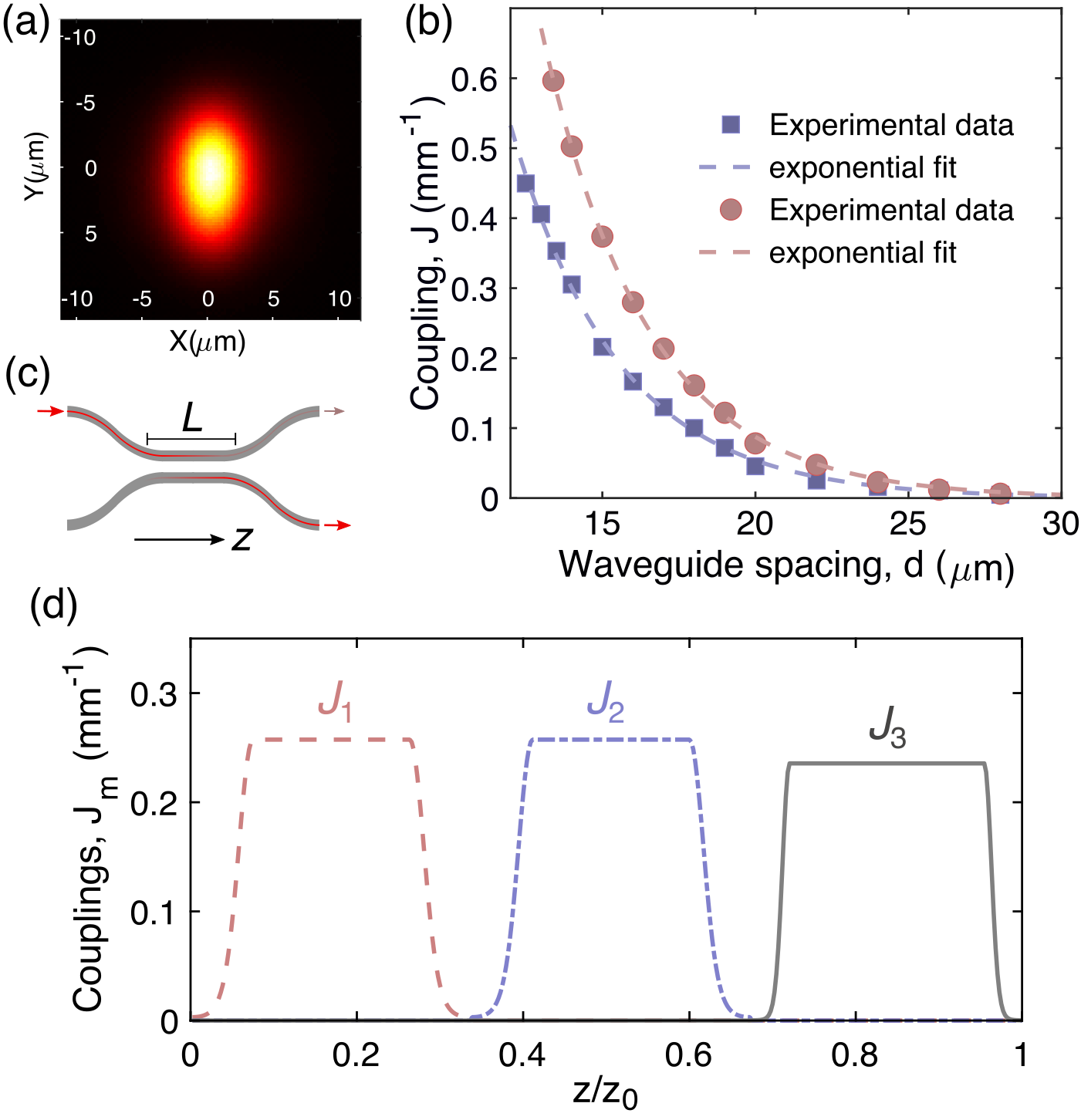}
\caption{(a) Intensity profile of the fundamental mode supported by each waveguide of the honeycomb lattice measured at $1030$ nm wavelength of vertically polarized light.
(b) Experimentally measured coupling strength as a function of inter-waveguide spacing, $d$. Filled square (circle) data points are for diagonal (vertical) couplings. 
(c) Schematic of a synchronously bent waveguide pair (building block of the honeycomb lattice) with a straight section of length $L$ where the coupling strength has a fixed value; see (d). 
For such a waveguide pair, $J(z)$ is estimated from the designed $d(z)$ as shown in (d) for the three couplings  $J_{1,2,3}$ in the lattice.
}
\label{threeCouplings}
\end{figure}

{\it Estimation of $\Lambda$.$-$} The parameter $\Lambda_m$ [$m\!=\!1,2,3$] for three driving steps was determined by characterizing a set of two-coupled waveguide systems. It should be highlighted that the fundamental mode supported by each waveguide in the lattice has an elliptical shape, Fig.~\ref{threeCouplings}(a). In the absence of circular symmetry of the mode, the 
coupling strength at a given wavelength of light is a function of the waveguide spacing $d$ as well as the polar angle $\theta$. In the case of a honeycomb lattice, $\theta\!=\!\{90^{\circ} , 30^{\circ}\}$ for the vertical and diagonal couplings, respectively. Fig.~\ref{threeCouplings}(b) shows the variation of inter-waveguide evanescent coupling as a function of $d$, for the  above-mentioned two polar angles -- filled square (circle) data points represent the diagonal (vertical) couplings.
The couplings vary exponentially with waveguide spacing. The designed values of the minimal waveguide spacing $d_{min}$ and straight coupling length $L$ for the topological lattice are $\{ d_{min}, L \}\!=\!\{ 16.5 \, \mu {\text{m}}, 5.15 \, {\text{mm}} \}$ and $\{ 14.5 \, \mu {\text{m}}, 4.15 \, {\text{mm}} \}$ for the vertical and diagonal couplings. 
One can estimate the couplings $J_m$ for the three driving steps from the known inter-waveguide spacing $d(z)$, see Fig.~\ref{threeCouplings}(d). 
The parameter $\Lambda_m\!=\! \int {\text{d}}z J_m(z)$ is then determined by integrating each $J_m$ over the driving step and was found to be $1.32$. To estimate the random fluctuation in $\Lambda$ due to fabrication imperfections, we created and characterized a set of thirteen synchronously bent two-waveguide devices (Fig.~\ref{threeCouplings}(c)), and the standard deviation in $\Lambda$ was found to be $0.05$.



{\it Probing topological edge states.$-$} The periodically driven honeycomb lattice with $\Lambda\!=\!1.32$ supports two quasienergy bands, each characterized by zero Chern number. However, as shown in Fig.~\ref{fig1}(c) in the main text, there exist 
chiral edge channels 
on both quasienergy gaps centered around $\varepsilon\!=\!\{0, \Omega/2\}$. Such a topological insulator, called an anomalous Floquet topological insulator~\cite{rudner2013anomalous}, is unique to periodically driven systems and is characterized by a distinct topological invariant known as the Floquet winding number. Unlike the Chern number, the winding number takes into account the full $z$-dependent propagator $\hat{U}(z)$ for the entire driving period and is equal to the number of edge channels in a quasienergy gap. 

\begin{figure}[]
\center
\includegraphics[width=\linewidth]{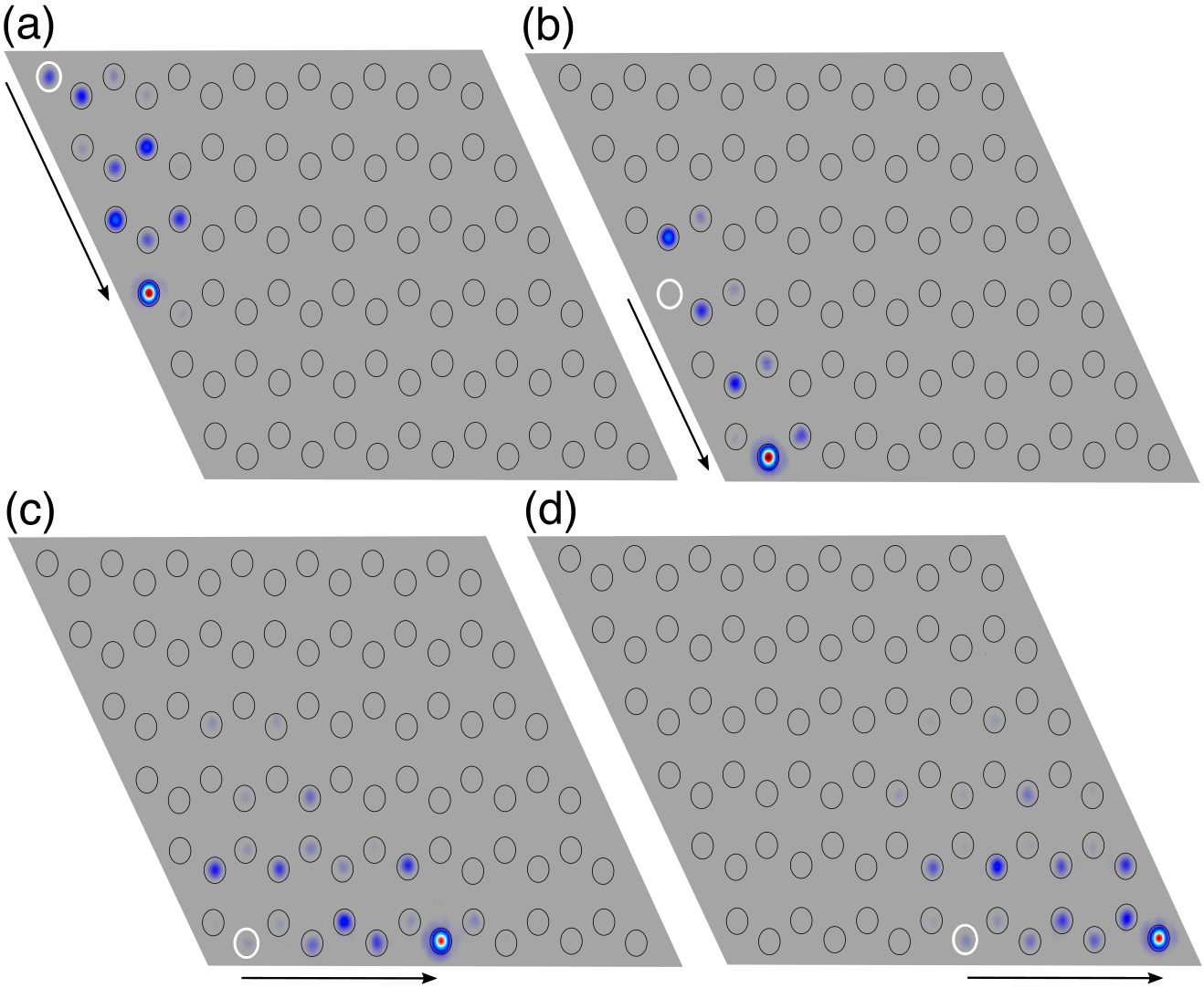}
\caption{Experimentally observed linear topological edge modes. To probe these edge modes, low power light was launched into a single edge waveguide at four different locations (a-d) indicated by the white circles on each image. The output intensity patterns were measured after a propagation of three complete driving periods, i.e., $3z_0$. In
these linear experiments, the single site input state efficiently overlaps with the topologically-protected edge modes and one-way edge transport of light was observed, demonstrating that the system is topological. }
\label{LinearEdgeStates}
\end{figure}

To demonstrate topologically robust edge transport, we launched light into single edge waveguides and measured the output intensity pattern after $z\!=\!3z_0$. 
In these measurements, the single site input state efficiently overlaps with the edge modes of the lattice -- the calculated overlap is $\approx\! 62\, \%$.
We observed backscatter-free unidirectional propagation of light traveling around corners, as shown in Figs.~\ref{LinearEdgeStates}(a-d). 

\begin{figure}[]
\center
\includegraphics[width=\linewidth]{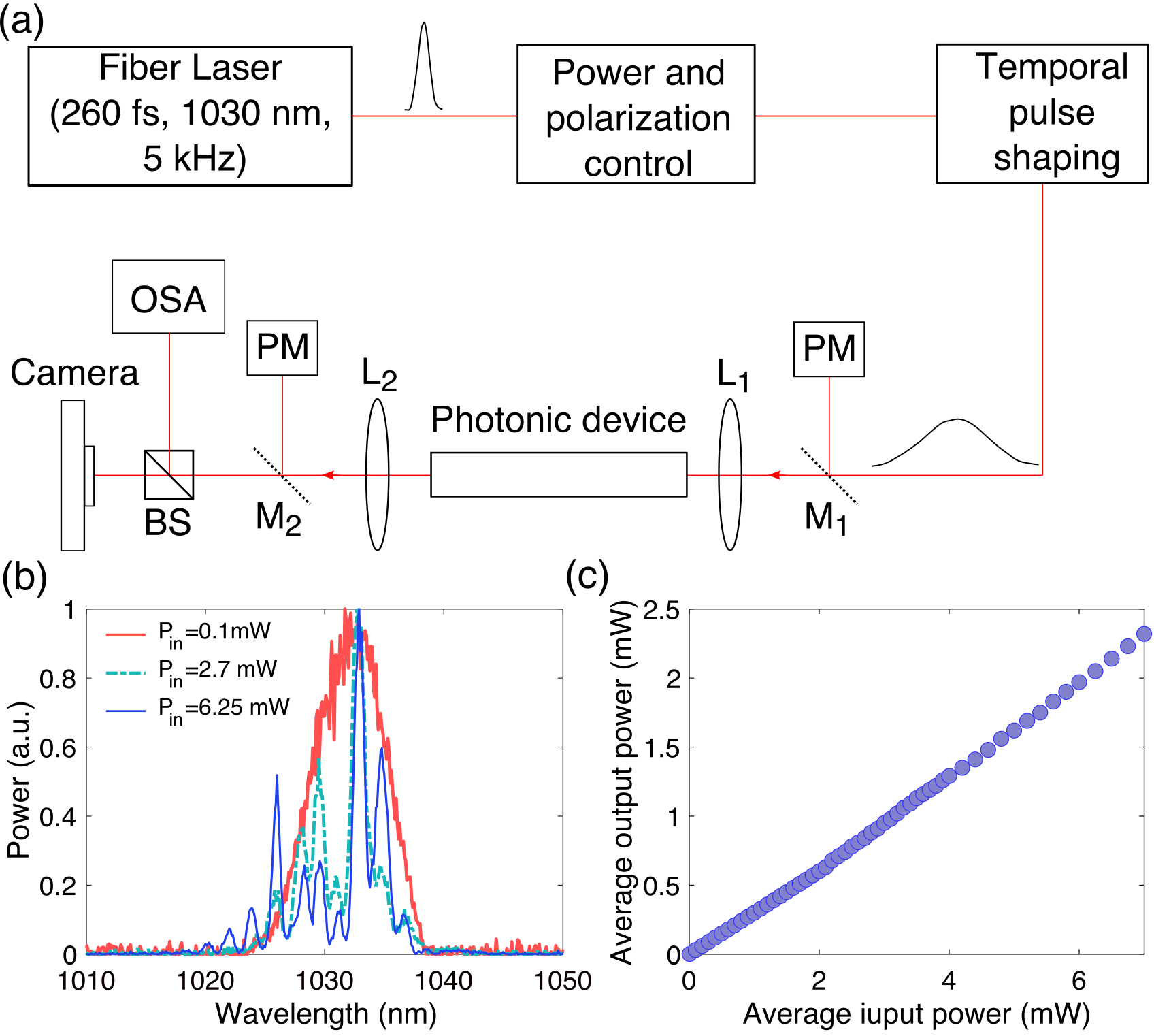}
\caption{Experimental setup for optical state preparation and probing nonlinear  states. The fiber laser generates $260$ fs pulses with a maximum pulse energy of $4 \, \mu$J.  
The laser power and polarization state is controlled using a system of wave plates and a polarizing beam splitter. Using a parallel grating pair, we down-chirp and stretch the pulses to $2$ ps. Here, PM: power meter, L$_{1,2}$: convex lenses, M$_{1,2}$: silver coated mirrors on flip mounts, BS: beam splitter, OSA: optical spectrum analyzer. (b) Output spectrum at $z\!=\!76$ mm for three different average input powers in mW. Spectrum broadening due to SPM is $\lesssim 20$ nm. (c) The output power varies linearly with input power for all measurements, indicating insignificant nonlinear losses.}
\label{setup}
\end{figure}

{\it Nonlinear characterization.$-$} For all nonlinear characterizations, we used the experimental setup shown in Fig.~\ref{setup}(a). The laser power and polarization state were controlled using a system of wave plates and a polarizing beam splitter. 
For the optical state preparation, we used a parallel grating pair to down-chirp and temporally stretch $260$ fs pulses to $2$ ps. 
For precise alignment, the in/out coupling lenses (L$_{1,2}$) were mounted on three-axis translation stages, and the photonic device was mounted on four-axis stages with two translational and two rotational degrees of freedom. The output intensity pattern was imaged on a CMOS camera. The output spectrum (intensity as a function of wavelength) was measured using an optical spectrum analyzer (OSA). 

The wave function $\psi_s$ in the discrete nonlinear Schr\"odinger equation, Eq.~\ref{eq1}, is a function of propagation distance. To experimentally access the nonlinear term, we use intense laser pulses; out of the continuous wave (CW) regime, one must think of the wavefunction at the $s$-th site is a function of both propagation distance and time $\psi_s(z)\leftrightarrow\psi_s(z, t)$. 
The nonlinear pulse propagation can in general be quite complex, and many undesired effects may arise such as chromatic dispersion, self-phase modulation (SPM) and multi-photon absorption. Given the group velocity dispersion and width of the laser pulses, we can safely neglect the effect of chromatic dispersion for a short propagation distance, $76$ mm. On the other hand, SPM (i.e., generation of new wavelengths) can not be neglected; however, it can be significantly reduced. To this end, we stretched and simultaneously down-chirped the laser pulses. In Fig.~\ref{setup}(c), the spectra at $z\!=\!76$ mm for three different input powers are shown. For low power, $P_{in}\!=\!0.1$ mW, the spectral width is $\approx \! 8 $ nm (FWHM). Due to the temporal pulse shaping, the maximal spectral width was observed to be $<20$ nm. 
Note that the evanescent couplings are a function of the wavelength of light and unless SPM is reduced, Eq.~\ref{eq1} can not be justified. For the spectral broadening observed in our experiments, the measured value of $\Delta J$ (i.e., variation of $J$ in the wavelength range of $1030\pm10$ nm) was $5.5\%$. The measured $\Delta J$ is of the order of disorder present in the waveguide lattice; hence, we can neglect the effect of SPM. In addition to this, we monitored both input and output power for all experiments to ensure that the nonlinear losses are insignificant~\ref{setup}(d).

\begin{figure}[t!]
\center
\includegraphics[width=\linewidth]{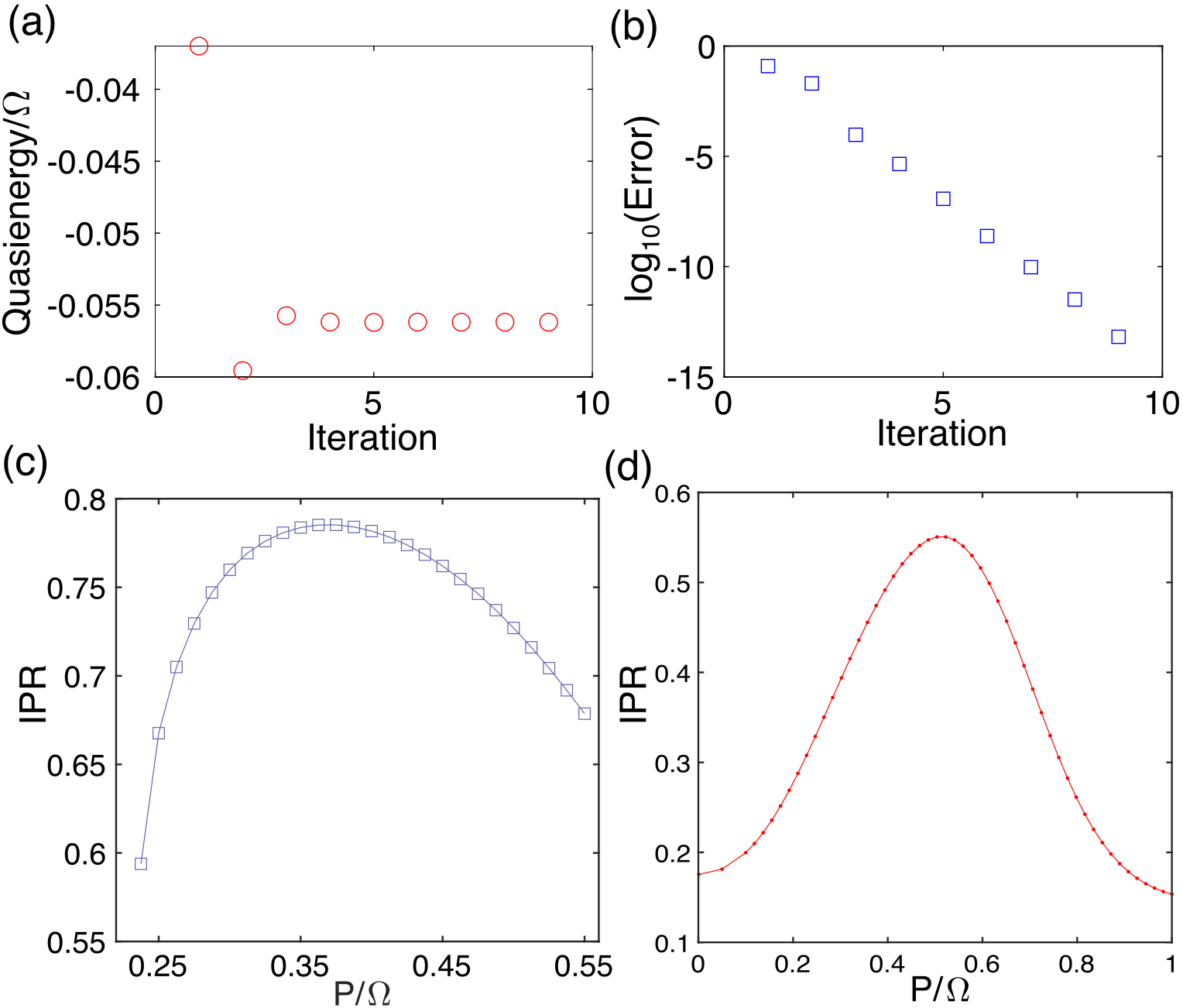}
\caption{
(a) Iteration process in a typical self-consistency algorithm. The convergence of quasienergy and (b) the error (defined as $\sum |\psi_{s}^{j}-\psi_{s}^{j+1}|^2/P$) calculated in each iteration step. The associated norm is $P\!=\!0.45\Omega$. (c) Size or spatial extent of period-doubled solitons. The IPR is a measure of localization. The soliton at the mid-gap quasienergy $\varepsilon\!=\!0$ (for which normalized power $P$ is near $0.35\Omega$) is most localized. (d) In experiments, this soliton family can be probed by injecting light into a single site because these Floquet solitons are strongly localized on a single site. Numerically calculated variation of IPR after a propagation of $3z_0$ for single site excitation. }
\label{iteration}
\end{figure}

{\it Self-consistency algorithm.$-$} In general, soliton solutions can be obtained using a numerical method known as the self-consistency algorithm~\cite{lumer2013self}. We modified this algorithm to obtain period-doubled Floquet solitons. 
The goal is to iteratively calculate self-localized nonlinear solutions of Eq.~\ref{eq1} starting from an appropriate initial guess state. 

When the evolution of a period-doubled Floquet soliton is considered, the total Hamiltonian of the system consists of a linear part $\hat{H}^{\text{lin}}$ and a nonlinear part $\hat{H}^{\text{nl}}$ arising due to the dynamical variation of the onsite energy (i.e., refractive index in our optical setup), $\hat{H}^{\text{tot}}\!=\!\hat{H}^{\text{lin}}+\hat{H}^{\text{nl}}$. The linear Hamiltonian of our anomalous Floquet topological insulator is $z$-periodic with a periodicity of $z_0\!=\! 2 \pi / \Omega$.  
The nonlinear part of the Hamiltonian, and hence the total effective linear Hamiltonian, has a periodicity of $2z_0$ because a period-doubled soliton reproduces itself (up to a phase factor) after an integer multiple of two driving periods. Using the self-consistency algorithm, we look for the solution of the total Floquet Hamiltonian $\hat{H}^{\text{tot}}(z)\!=\!\hat{H}^{\text{tot}}(z+2z_0)$.  Here, we consider the potential generated by the wavefunction through the nonlinear term as a linear potential. It should be noted that these nonlinear solutions are allowed to have `micromotion,' meaning they can change shape within their periodic motion.

We consider a tight-binding honeycomb lattice of $288$ sites with periodic boundary condition. The nearest-neighbor couplings $J_{1,2,3}$ are modulated in a three-step clockwise manner, as explained in the main text. All higher-order long-range couplings are assumed to be negligible. 
The initial guess state $\psi_{{s}}^{j}(z\!=\!0)$ [$j$ indicates iteration step] is taken to be localized on a single bulk site $s=\tilde {s}$. For a given norm $P=\sum |\psi|^2$, the guess state is propagated from $z\!=0$ to $2z_0$ using Eq.~\ref{eq1} to obtain $\psi_{s}^{j}(z)$. We treat $\psi_{s}^{j}(z)$ as a periodic function of $z$ such that the nonlinear part of the Hamiltonian associated with Eq.~\ref{eq1} is $z$-periodic with a period of $2z_0$. We then use 
Floquet theory to evaluate the eigenvalues (quasienergy) and eigenfunctions of the total Hamiltonian. 
We pick the eigenfunction $\psi_{s}^{j+1}$ that maximally overlaps with the normalized initial guess state.
In the next step of the iteration, the correctly normalized $\psi_{s}^{j+1}$ is taken to be the new
initial state. We continue the iteration process until
the quasienergy exhibits convergence and the error 
(defined as $\sum |\psi_{s}^{j}-\psi_{s}^{j+1}|^2/P$) 
is less than a threshold that is sufficiently small (typically $10^{-12}$), see Figs.~\ref{iteration}(a,b).


\begin{figure}[t!]
\center
\includegraphics[width=\linewidth]{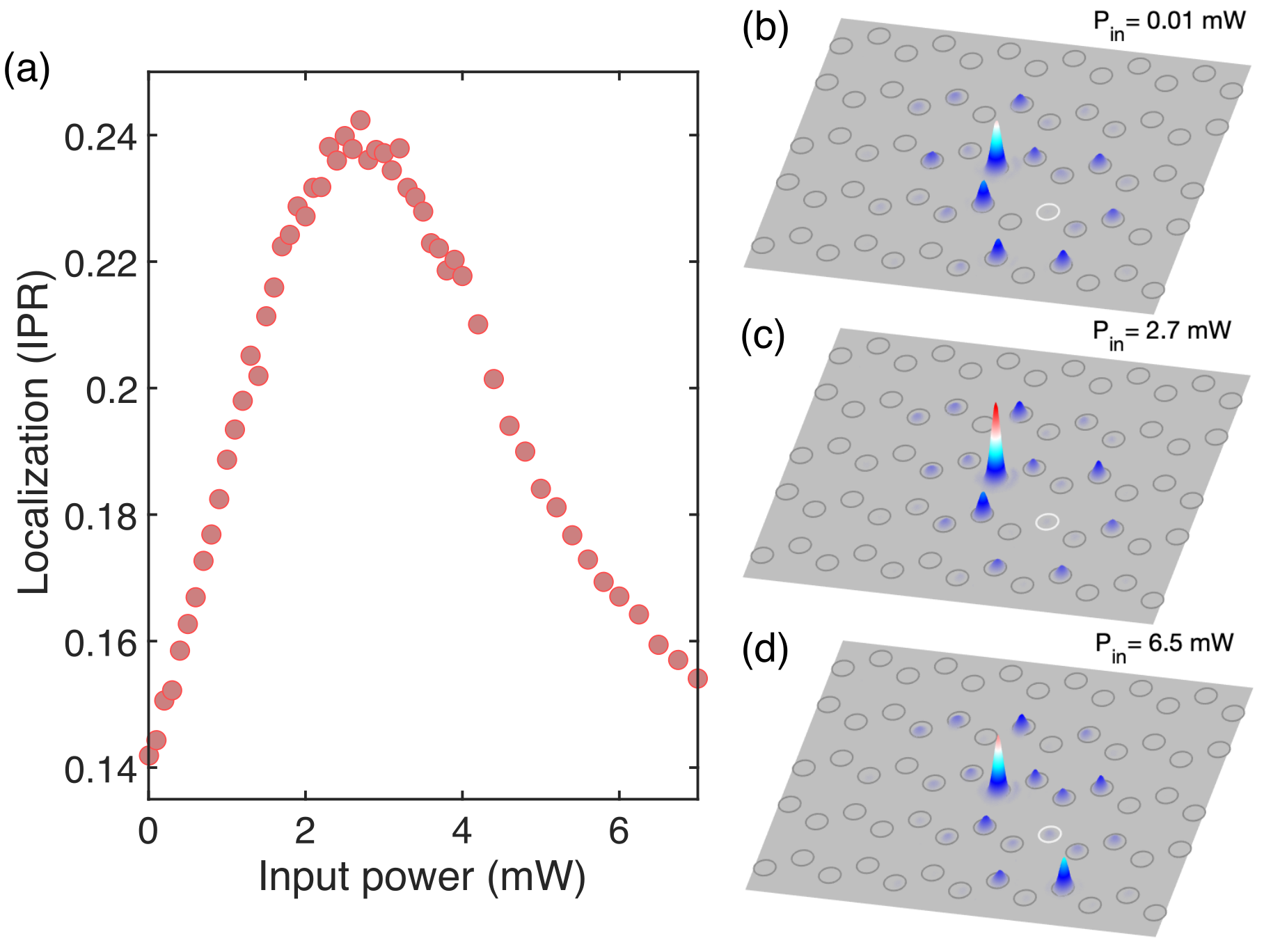}
\caption{(a) Experimentally measured variation of the inverse participation ratio (IPR) after a propagation of $3z_0$ as a function of average input power. The observed peak in the IPR confirms the existence of the family of Floquet solitons. (b-d) Output intensity patterns for three different input powers that are indicated on each image. (c) is the same as Fig.~\ref{fig3}(f)}.
\label{IPR}
\end{figure}

\begin{figure}[t!]
\center
\includegraphics[width=\linewidth]{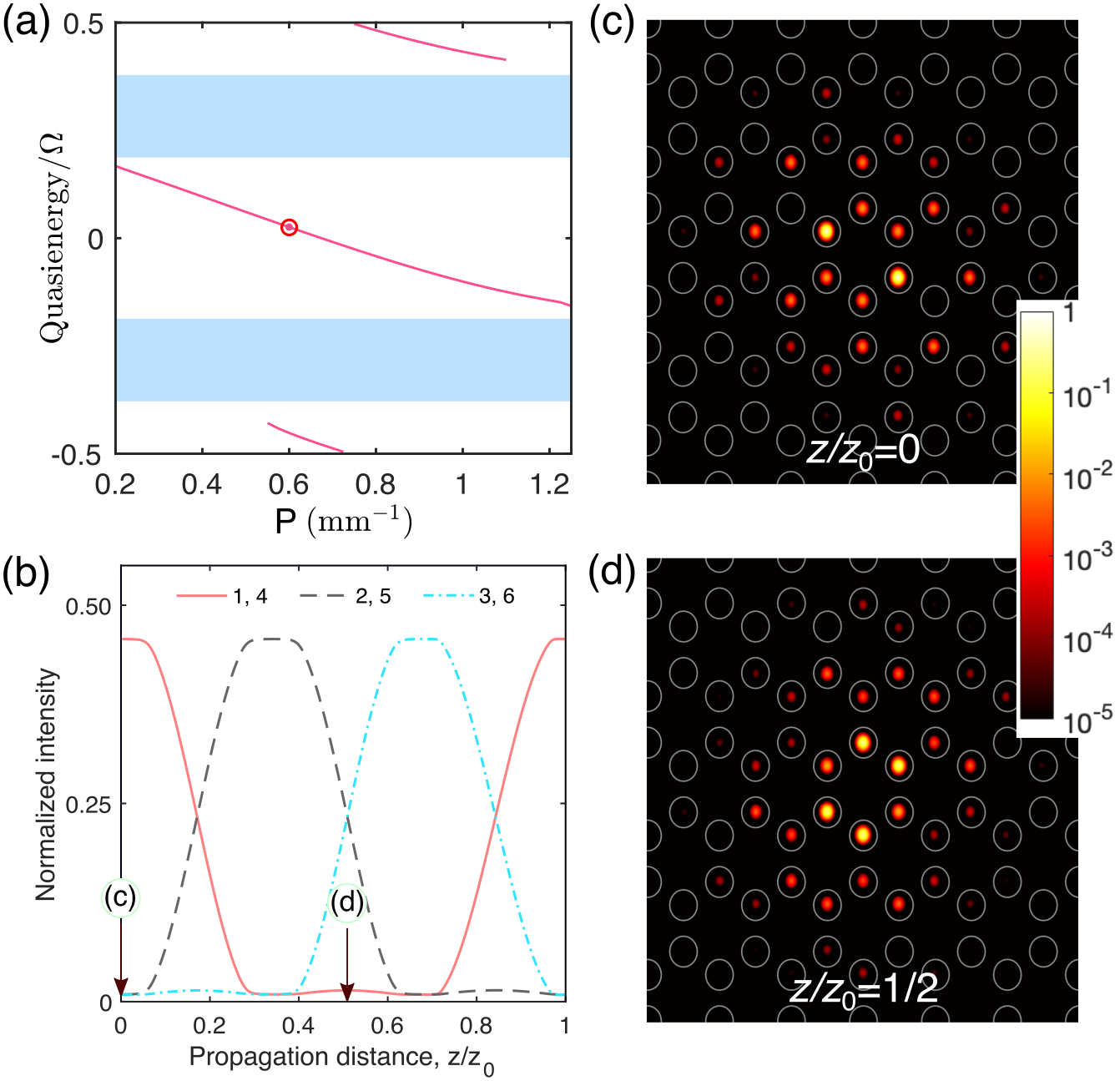}
\caption{(a) Quasienergy spectrum as a function of renormalized power showing linear bulk modes (blue) and a family of ``regular" Floquet solitons (red) that have the same period as the driving.
Here $\Lambda\!=\!0.42\pi$. 
(b) Variation of normalized intensity at the six sites (see Fig.~\ref{fig2}e) showing the dynamics of the soliton for a complete driving periods.
(c,d) Intensity distributions of a ``regular" Floquet soliton at a renormalized power $P\!=\!0.6 \Omega$ [indicated by the red circle in (a)] for $z/z_0\!=\!0$ and $1/2$, respectively. Note that the soliton exhibits a cyclotron-type rotation. }
\label{singlePeriod_Sol}
\end{figure}

\begin{figure}[b!]
\center
\includegraphics[width=\linewidth]{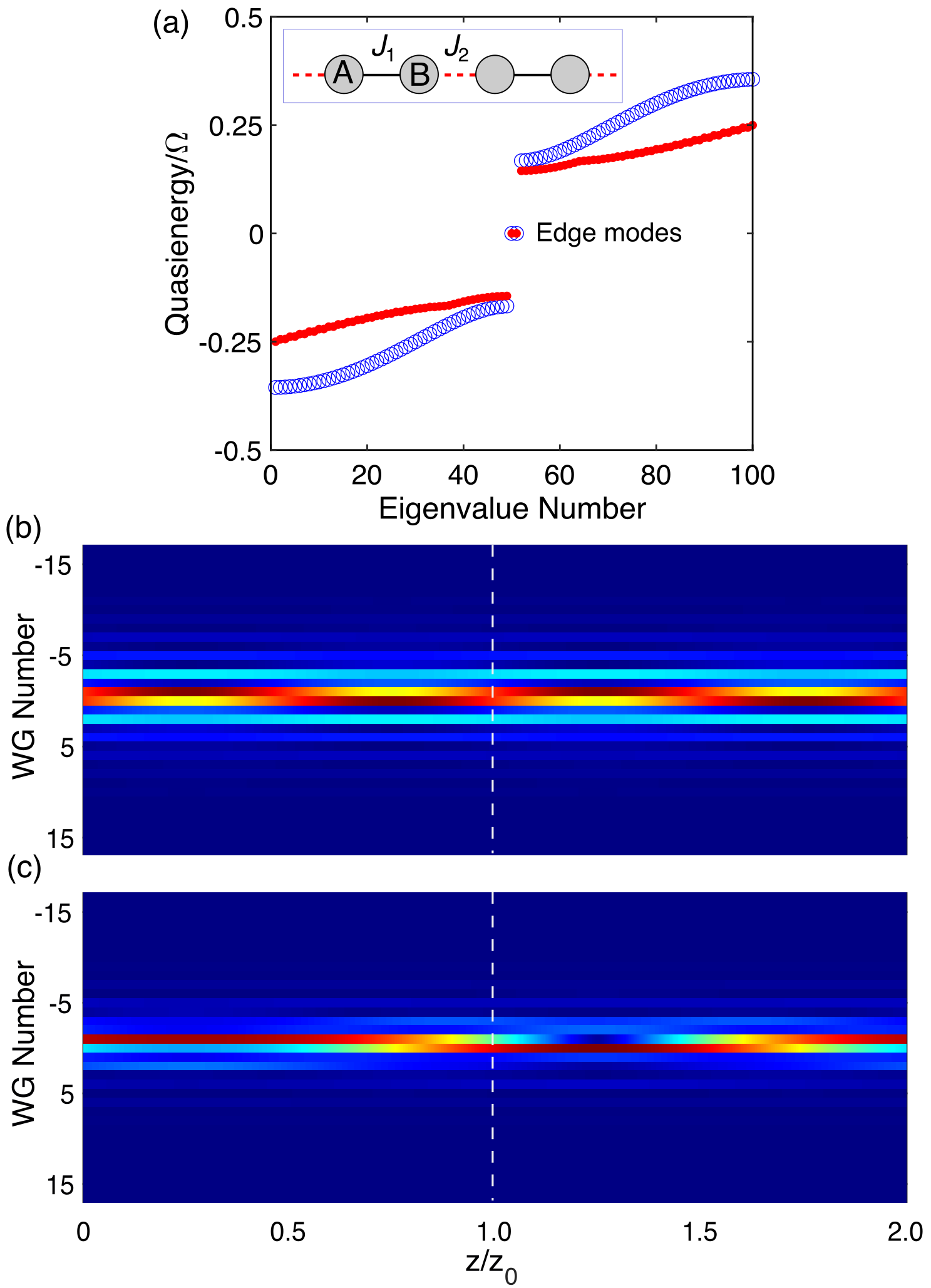}
\caption{(a) Quasienergy spectrum of a sinusoidally driven SSH lattice of $100$ sites (blue). The red data corresponds to period-doubled spectrum, i.e., after band folding. Note the existence of the quasienrgy gap around $\varepsilon\!=\!0$ after band folding. 
The inset shows the intra- and inter-cell hoppings.
(b) Dynamics of a traditional single-period Floquet soliton residing in the topological band gap. (c) Dynamics of a period-doubled Floquet soliton that reproduces itself after two driving periods. The white dashed line indicates $z/z_0\!=\!1$.}
\label{drivenSSH}
\end{figure}

{\it Soliton size.$-$} As mentioned in the main text, Floquet solitons show fundamentally different localization features compared to static solitons. 
%
For example, as the strength of the focusing nonlinearity increases, a soliton in an undriven square lattice becomes increasingly localized because spectrally it moves away from the linear modes. 
On the other hand, the family of period-doubled Floquet solitons resides in a quasienergy gap surrounded by linear modes from both sides, see Fig.~\ref{fig2}(a). As a result, when the nonlinearity is increased, we observe crossovers from delocalization to localization to delocalization. In other words, the mid-gap soliton near quasienergy $\varepsilon\!=\!0$ is maximally localized. In Fig.~\ref{iteration}(c), we plot the inverse participation ratio (IPR) of the soliton wave function, and a clear IPR peak is observed. 
Observation of such an IPR peak is experimental proof of the existence of the family of solitons. Note that all members of the soliton family are strongly peaked on a single lattice site; hence, one can launch light into a single site with a tunable optical power to probe these Floquet solitons. Fig.~\ref{iteration}(d) shows the numerically calculated variation of IPR as a function of nonlinearity: a clear IPR peak is observed. Additionally, the variation of IPR in Fig.~\ref{iteration}(d) is quantitatively different from Fig.~\ref{iteration}(c) because the initial state is a single site excitation rather than a soliton. 
In this context it should be noted that the stroboscopically calculated IPR for the soliton family is independent of the propagation distance. In other words, if exact soliton solutions are launched experimentally, we would observe Fig.~\ref{iteration}(c) after each integer value of the driving period. On the other hand, for a single-site input state, the IPR is a function of propagation distance. In general the variation of IPR with $P$ can be complicated. However, in the limit of weakly dispersive bulk bands, a clear IPR peak can be observed overage a wide range of propagation distances.

Fig.~\ref{IPR}(a) shows the experimentally measured variation of IPR after a propagation distance of $3z_0$, and the corresponding intensity patterns at three different input powers are shown in Fig.~\ref{IPR}(b-d). At low power, we see significant diffraction. As the optical power is increased, the output intensity pattern becomes more and more localized. Near $2.7$ mW, the maximally localized output state, associated with the IPR peak, is observed. At higher power, the output state becomes delocalized again as the soliton approaches the other band edge, as would be expected from numerical calculations.

{\it Single-period solitons.$-$} Thus far, we have only discussed period-doubled Floquet solitons in the honeycomb lattice. The same lattice also supports the traditional Floquet solitons, whose periodicity is the same as the driving period $z_0$. We obtained families of such single period Floquet solitons residing in both topological quasienergy gaps, using the self-consistency algorithm, see Fig.~\ref{singlePeriod_Sol}(a). Fig.~\ref{singlePeriod_Sol}(b) shows the dynamics of a single period soliton, and the intensity patterns at two different propagation distances are presented in Fig.~\ref{singlePeriod_Sol}(c,d). Note that the single period soliton has two equally intense peaks, and it exhibits a clockwise rotational motion similar to the period-doubled solition.  This single period soliton can be thought of roughly as consisting of two period-doubled solitons that rotate about one another.

{\it Period-doubled solitons in a driven SSH lattice.$-$} 
The Su-Schrieffer-Heeger (SSH) model~\cite{su1979solitons} is a one-dimensional lattice model with two sites ($A$ and $B$) per unit cell. The intra- and inter-cell hoppings are denoted by $J_{1,2}$, respectively; see the inset of Fig.~\ref{drivenSSH}(a).
In this section, we consider a sinusoidally-modulated nonlinear SSH lattice and numerically calculate period-doubled solitons.  In our experiments, such a driven lattice can be realized by sinusoidally modulating the 
paths 
of the waveguides in the array. 
In the scalar-paraxial tight-binding approximation, the dynamics of light along the driven SSH lattice is governed by 
\begin{eqnarray}
\label{ssh}
i\frac{\partial}{\partial z} \psi^A_s\!=\!&
\!-\! J_1\psi^B_{s-1} \!-\! J_2\psi^B_{s} 
\!-\! s V(z) \psi^A_{s} \!-\! |\psi^A_s|^2 \psi^A_s, \nonumber  \quad \quad \\
i\frac{\partial}{\partial z}  \psi^B_{s}\!=\!&
\!-\! J_1\psi^A_{s} \!-\! J_2\psi^A_{s+1} 
\!-\! (s\!+\!\frac{1}{2}) V(z) \psi^B_{s} \!-\! |\psi^B_s|^2 \psi^B_s, \nonumber \\
\end{eqnarray} 
where $V(z)\!=\!K\sin(\Omega z)$, $s$ enumerates the unit cells, $K$ is the driving amplitude and $\Omega\!=\!2\pi/z_0$ is the driving frequency. 
Here the modulation $V(z)$ gives the periodic modification of the lattice potential, and we use it 
to engineer the band structure. 
We note that such modulation of the waveguide positions as a function of $z$ is equivalent to applying a sinusoidal electric field along the axis of a one-dimensional lattice of electrons~\cite{longhi2006observation, szameit2010observation, mukherjee2015modulation}.
In numerical calculations, we consider the following parameters: $J_2/J_1\!=\!0.35$, $\Omega\!=\!3.5J_1$, $K=0.25\Omega$. We have chosen this set of parameters to ensure that a quasienrgy gap exists around $\varepsilon\!=\!0$ after the band folding -- the criterion for realizing period-doubled solitons; see main text. The blue data points in Fig.~\ref{drivenSSH}(a) present the quasienergies for a finite lattice of $100$ sites. The red data sets correspond to the quasienergies after band folding.

Using the self-consistency algorithm, we first look for the traditional single-period solitons that bifurcate from the `upper' quasienergy band. The dynamics of such a soliton at $P=1.4$ is shown in Fig.~\ref{drivenSSH}(b) for two complete periods. The traditional Floquet soliton repeats itself after each driving period, up to a phase factor.
We then look for period-doubled solitons in the same lattice; in other words, nonlinear solutions of Eq.~\ref{ssh} that repeat themselves after two driving periods.
In Fig.~\ref{drivenSSH}(c), we show the dynamics of such a period-doubled soliton at $P=1.4$. 

In the main text, we demonstrate a period-doubled soliton using a specific driving -- a honeycomb lattice with a three-step driving protocol. In this section, we use a driven SSH lattice to show that period-doubled solitions may be found as long as a quasienergy gap exists after band folding; this suggests that period doubled solitons are a general phenomnon that may be found in a variety of gapped driven systems.

\clearpage
\end{document}